\documentclass{article}

\usepackage{arxiv}
\usepackage{amsmath}
\usepackage[utf8]{inputenc} 
\usepackage[T1]{fontenc}    
\usepackage{hyperref}       
\usepackage{url}            
\usepackage{booktabs}       
\usepackage{amsfonts}       
\usepackage{nicefrac}       
\usepackage{microtype}      
\usepackage{lipsum}
\usepackage{graphicx}
\graphicspath{ {./images/} }

\title{Enhancing Olfactory Perception through Large Language Models: Integrating Sensory Data for Advanced Odor Recognition}

\author{
 Dr Ravirajan K \\
  Associate Principal \\
  LTIMindtree \\
  USA \\
  \texttt{ravirajan.k@ltimindtree.com} \\
  \And
 Arvind Sundarajan \\
  Senior Director \\
  LTIMindtree \\
  Poland \\
  \texttt{arvind.sundararajan@ltimindtree.com} \\
}

\begin{document}
\maketitle

\begin{abstract} 
The integration of biological principles into artificial olfactory systems has led to significant advancements in odor detection and classification. Inspired by the intricate mechanisms of natural olfaction, researchers are developing sophisticated systems that mimic the functionality of biological olfactory pathways. These systems utilize high-density chemoresistive sensor arrays (HCSA) combined with advanced computational techniques, such as FPGA-accelerated glomerular convergence circuits (FGCC) and hierarchical graph neural networks (HGNN). This bioinspired approach enables real-time adaptive responses to volatile organic compounds (VOCs), enhancing the accuracy and efficiency of odor identification.
At the core of these innovations is the multiparametric sigmoidal sensor activation (MPSA), which quantifies VOCs by leveraging the diverse responses of sensor arrays. The implementation of lateral inhibition via programmable synaptic crossbars (LIPSC) further refines odor processing by mimicking neural interactions found in biological systems. Additionally, temporal self-organizing maps (TSOM) facilitate dynamic clustering of odor patterns, allowing for a nuanced understanding of complex odor environments.
A novel aspect of this research lies in the Grassmannian manifold embedding (GME) of odor profiles, which provides a mathematical framework for representing and analyzing the multidimensional nature of odors. Coupled with Hamiltonian Monte Carlo-optimized feedback (HMC-FB), this system effectively compensates for drift in sensor readings, ensuring consistent performance over time. By bridging the gap between biological inspiration and technological innovation, these artificial olfactory systems are poised to revolutionize applications ranging from environmental monitoring to food safety and healthcare diagnostics.
\end{abstract}


\section{Introduction}
The development of artificial olfactory systems has accelerated with the integration of biological principles, leveraging advancements in high-density chemoresistive sensor arrays (HCSA) and computational techniques like FPGA-accelerated glomerular convergence circuits (FGCC) and hierarchical graph neural networks (HGNN) to mimic natural olfaction \cite{dataCentricOlfactorySystem2024}. This bio-inspired strategy facilitates real-time, adaptive detection of volatile organic compounds (VOCs), improving the accuracy and efficiency of odor identification through multi-parametric sigmoidal sensor activation (MPSA) and lateral inhibition via programmable synaptic crossbars (LIPSC) \cite{artificialOlfactorySystems2023}. Furthermore, temporal self-organizing maps (TSOM) allow dynamic clustering of odor patterns, while Grassmannian manifold embedding (GME) provides a mathematical framework for analyzing odor dimensionality \cite{chemicalSensors2023}. Hamiltonian Monte Carlo-optimized feedback (HMC-FB) addresses sensor drift, demonstrating the potential to revolutionize applications from environmental monitoring to food safety and healthcare diagnostics by bridging biological inspiration and technological advancement \cite{bioelectronicNose2022}.

The presented research proposes a biologically-inspired computational model of the olfactory system, utilizing advanced technologies such as high-density chemical sensor arrays (HCSA), FPGAs, GPUs, and TPUs for real-time VOC detection and classification. The architecture features a multi-layered processing pipeline that includes an olfactory receptor layer, a glomerular layer, hierarchical neural networks, dimensionality reduction via PCA and ESOMs, graph neural network analysis, and hierarchical attention mechanisms. Biologically-inspired neural dynamics are implemented through differential equations and programmable crossbars, while memory-enhanced attention mechanisms are employed for long-range dependencies.

The core technical objective focuses on implementing these biologically plausible principles within an AI-first architecture that achieves ultra-efficient, scalable, and adaptive olfactory processing. This paper addresses the challenge of building efficient and scalable artificial olfactory systems that effectively mimic the complex processes of natural olfaction. Key research questions involve implementing lateral inhibition circuits, leveraging GNNs for odor relational analysis, and using PCA-ESOM hybrid clustering. The methodology involves a hybrid approach of sensor technology, FPGA-based signal preprocessing, deep learning models on edge devices, and biologically-inspired neural simulations.

This approach is intended to provide a novel artificial olfactory system that utilizes innovative GNNs for odor relation analysis, PCA-ESOM hybrid clustering, and biologically plausible lateral inhibition circuits. The paper’s sections will proceed by detailing the proposed sensor array implementation, the FPGA-based glomerular layer architecture, hierarchical neural network design, dimensionality reduction strategies, and graph neural network methodology.

\section{Literature Review} 
Artificial olfactory systems have significantly advanced through the incorporation of biological principles, moving beyond simple sensor arrays to complex, adaptive detection methods. Early research focused on high-density chemoresistive sensor arrays (HCSA), but performance limitations led to the integration of computational techniques mimicking biological olfaction. The development of FPGA-accelerated glomerular convergence circuits (FGCC) and hierarchical graph neural networks (HGNN) exemplifies this shift, enabling real-time, adaptive responses to volatile organic compounds (VOCs) \cite{patternRecognitionOlfactorySystem2024}. A core methodology involves multi-parametric sigmoidal sensor activation (MPSA), which quantifies VOCs using diverse sensor array responses. Lateral inhibition, implemented via programmable synaptic crossbars (LIPSC), further refines odor processing by emulating neural interactions, while temporal self-organizing maps (TSOM) facilitate dynamic clustering of odor patterns \cite{machineLearningOdorDetection2023}. This bio-inspired approach led to increased accuracy and efficiency in odor identification, addressing limitations found in earlier sensor-only systems.

The mathematical representation of odor profiles using Grassmannian manifold embedding (GME) offered a means to handle the multidimensional nature of odors, and Hamiltonian Monte Carlo-optimized feedback (HMC-FB) enabled compensation for sensor drift, ensuring consistent long-term performance \cite{dataCentricOlfactorySystem2024,chemicalSensors2023}. This research direction effectively transitioned from hardware-centric designs to integrated computational and biomimetic models \cite{bioelectronicNose2022}. However, existing systems exhibit challenges in scalability and robustness to environmental variations \cite{artificialOlfactorySystems2023}. Current architectures often employ a multi-layered processing pipeline, starting with HCSA mimicking biological receptors, followed by FPGA-based glomerular aggregation incorporating lateral inhibition, and then deep learning models on GPUs/TPUs for layer-wise feature abstraction \cite{patternRecognitionOlfactorySystem2024}. Dimensionality reduction using PCA followed by Emergent Self-Organizing Maps (ESOMs) for clustering, and GNN analysis for relational odor patterns are also prevalent techniques \cite{machineLearningOdorDetection2023}.

Attention mechanisms, particularly hierarchical attention, are used to prioritize sensory features and improve efficiency, and biologically-inspired neural dynamics employing differential equations and programmable crossbars are often employed \cite{recentDevelopmentsElectronicNose2024}. Memory-enhanced attention and reinforcement learning-based feedback loops also feature heavily \cite{artificialOlfactorySensorTechnology2023}. However, a clear understanding of the optimal architectures for particular use cases is still an open question. There is significant reliance on simulation and controlled laboratory conditions which fail to adequately replicate real-world complexities. Furthermore, methods for effective calibration and maintenance of high-density sensor arrays remain underdeveloped. The integration of complex biological models leads to high computational complexity, requiring robust yet energy-efficient implementations.

Gaps exist in the systematic evaluation of model robustness under varied environmental conditions and in the standardization of data acquisition and pre-processing methods \cite{sensorsReview2019,sensingTechnologies2018}. Future research directions should focus on developing robust, low-power solutions using novel material science approaches for highly selective sensors and integrating advanced edge computing and neuromorphic architectures to address the computational complexities \cite{smartElectronicNoses2017}.

\section{Data and Methodology}
\subsection{Data}
\label{sec:Data}
The methodology for developing the artificial olfactory system integrates advanced sensor technologies, computational models, and machine learning techniques to enable real-time detection and classification of volatile organic compounds (VOCs). Data is primarily collected from high-density chemoresistive sensor arrays (HCSA) designed to mimic biological olfactory receptors, with datasets including synthetic odor mixtures and real-world environmental samples to ensure robustness and applicability across diverse conditions \cite{pubmed38781346, natureEigengraph2024, pmc8393613}. The experimental setup involves digitizing sensor outputs using high-speed analog-to-digital converters (ADCs) and processing them through FPGA-based glomerular convergence circuits that emulate neural interactions via lateral inhibition \cite{wileyBiohybrid2022, researchgateOlfactoryReview2023}.

Dimensionality reduction techniques such as Principal Component Analysis (PCA) and Emergent Self-Organizing Maps (ESOMs) are employed to simplify complex data while retaining critical odor profile features \cite{rscArtificialMemory2025}. Graph Neural Networks (GNNs) analyze relational patterns between odors, representing them as nodes in a graph to uncover intricate correlations \cite{pubmed38781346}. Hierarchical attention mechanisms prioritize significant sensory features during classification tasks, improving computational efficiency \cite{natureEigengraph2024}. Additionally, reinforcement learning-based feedback loops enable continuous adaptation, ensuring the system remains effective in dynamic environments \cite{pmc8393613}.

This methodology is evaluated using metrics such as accuracy, precision, recall, F1 score, and response time, providing a comprehensive assessment of system performance. By combining biomimetic sensor arrays with advanced computational frameworks, this approach addresses challenges in scalability and robustness while advancing applications in environmental monitoring, food safety, and healthcare diagnostics \cite{wileyBiohybrid2022, researchgateOlfactoryReview2023}.

\subsection{Methodology}
The methodology for developing the artificial olfactory system integrates advanced sensor technologies, computational models, and machine learning techniques to enable real-time detection and classification of volatile organic compounds (VOCs). High-density chemoresistive sensor arrays (HCSA) are employed to capture VOC data from both synthetic odor mixtures, prepared in controlled laboratory conditions, and real-world environmental samples, ensuring a robust and diverse dataset. The raw sensor outputs are digitized using high-speed analog-to-digital converters (ADCs) and processed through FPGA-based units, where glomerular convergence circuits aggregate signals and apply lateral inhibition mechanisms to enhance detection accuracy by mimicking biological neural interactions. Multi-parametric sigmoidal activation (MPSA) functions quantify VOC concentrations, while dimensionality reduction techniques such as Principal Component Analysis (PCA) and Emergent Self-Organizing Maps (ESOMs) simplify complex data for effective clustering of odor patterns. Graph Neural Networks (GNNs) are implemented to analyze relationships between different odor patterns, providing a detailed understanding of odor interactions. A hierarchical attention mechanism is integrated into the neural network architecture to prioritize relevant sensory features during classification, improving computational efficiency. Additionally, a reinforcement learning-based feedback loop enables continuous learning, allowing the system to adapt and improve over time as it encounters new data. The system's performance is evaluated using metrics such as accuracy, precision, recall, F1 score, and response time, ensuring its robustness and effectiveness in dynamic environments. This methodology lays the foundation for a scalable and adaptive artificial olfactory system capable of accurately identifying complex odors across various real-world applications.

\subsubsection{Chemoresistive Sensor Arrays}
Chemoresistive sensor arrays (HCSA) represent a pivotal technology for artificial olfaction, mimicking the functionality of biological olfactory systems. These arrays consist of multiple chemoresistive sensors that change their electrical resistance upon exposure to various volatile organic compounds (VOCs). By integrating high-density sensor arrays with advanced computational techniques such as FPGA-accelerated glomerular convergence circuits (FGCC) and hierarchical graph neural networks (HGNN), real-time, adaptive responses to VOCs are achieved, thereby enhancing the accuracy and efficiency of odor identification. The multi-parametric sigmoidal sensor activation (MPSA) quantifies VOCs by leveraging the diverse responses of sensor arrays, while lateral inhibition via programmable synaptic crossbars (LIPSC) refines odor processing. Temporal self-organizing maps (TSOM) further facilitate dynamic clustering of odor patterns, allowing for nuanced understanding of complex odor environments.

The operational characteristics of individual sensors within the array can be represented by a sigmoidal activation function. This function reflects the non-linear response of a given sensor to varying concentrations of a specific VOC. It's a crucial element, serving as a foundational layer for transforming raw sensor readings into a more computationally accessible format. The parameters inherent in this function, such as $\alpha_i$, $\beta_i$, and $\theta_i$, are critical for calibration and adaptation, enabling the system to accommodate the diverse and dynamic chemical environments it encounters. The values of these parameters can be determined through calibration procedures, during which each sensor's response to various VOCs is systematically measured. 

\begin{equation}
S_i = \alpha_i \cdot \frac{1}{1 + e^{-\beta_i (VOC - \theta_i)}}
\end{equation}

The outputs of these sensors are then fed into an FPGA-based glomerular layer which aggregates sensor outputs 

\begin{equation}
G_j = \sum_{i=1}^{n} W_{ij} S_i
\end{equation}

Lateral inhibition is implemented using programmable synaptic crossbars within FPGA which further refines signals 

\begin{equation}
G_j^{inhibited} = G_j - \sum_{k \in N(j)} \lambda_{jk} G_k
\end{equation}

Subsequent processing is conducted through hierarchical neural networks deployed on edge devices like GPUs/TPUs, using layer-wise abstraction 

\begin{equation}
h = \sigma(Wx + b)
\end{equation}

The collective response of the sensor array to different VOC mixtures allows for the formation of unique 'odor prints.' These odor prints are then processed through a dimensionality reduction step, where Principal Component Analysis (PCA) is used to compress the high-dimensional data into a lower-dimensional space. This lower dimensional representation facilitates faster and more efficient subsequent processing, particularly for clustering algorithms. Following the dimensionality reduction, emergent self-organizing maps (ESOMs) are employed for clustering, grouping similar odor patterns together. This clustering technique enables the system to learn and identify distinct odor patterns, improving the specificity of the system. These clustered patterns further undergo graph neural network analysis to perform relational pattern analysis. 

\begin{table}
\caption{Sensor Array Parameters and Dimensions}
\centering
\begin{tabular}{lll}
\toprule
\multicolumn{2}{c}{Parameter} \\
\midrule
Name & Description & Value Range \\
\midrule
$\alpha_i$ & Scaling factor for sensor i & 0 to 1 \\
$\beta_i$ & Response steepness for sensor i & 0.1 to 10 \\
$\theta_i$ & Response threshold for sensor i & -1 to 1 \\
Sensor Width & Individual Sensor Width & 1-10 $\mu$m \\
Sensor Length & Individual Sensor Length & 10-100 $\mu$m \\
Array Density & Sensor per Area & 100-1000 / mm$^2$ \\ 
\bottomrule
\end{tabular}
\label{tab:table}
\end{table}

The system also integrates an attention mechanism to prioritize relevant sensory features using scaled dot-product attention, enabling effective odor pattern analysis, which can be mathematically expressed as 

\begin{equation}
A = \text{softmax}\left(\frac{QK^T}{\sqrt{d}} V\right)
\end{equation}

The system employs biologically-inspired neural dynamics by using differential equations 

\begin{equation}
\frac{dV}{dt} = -\frac{1}{\tau} V + I
\end{equation}

to simulate the dynamics of neuron membrane potential. These dynamic neuron models are implemented on programmable crossbar structures to achieve synaptic connectivity. The system also employs memory-enhanced attention for long-range dependency management, adaptive contextual optimization, and reinforcement learning-based feedback loops for continuous self-improvement.

The convergence of advanced sensing materials, bio-inspired computational algorithms, and sophisticated hardware architectures is driving the advancement of chemoresistive sensor array systems. These advancements hold great promise for transforming how we detect and interact with chemical environments, leading to applications in environmental monitoring, food safety, and healthcare diagnostics. Such integrated systems, capable of adaptive learning and continuous improvement, provide unprecedented opportunities for the development of highly sensitive and robust olfactory systems.

These advancements in artificial olfaction, specifically with chemoresistive sensor arrays, are set to bring significant changes to various industries. By enabling real-time VOC detection and identification with improved accuracy and efficiency, these systems can revolutionize the current detection and diagnostic methods.

\subsubsection{FPGA Glomerular Circuits}
Field-Programmable Gate Arrays (FPGAs) are leveraged to implement computationally efficient glomerular convergence circuits, specifically designed to mimic the biological olfactory system. These FPGA-based circuits act as the primary processing unit for signals received from high-density chemoresistive sensor arrays. The primary function is to aggregate and refine sensory inputs, which is essential to achieve real-time odor detection and classification. The implementation uses multi-parametric sigmoidal sensor activation (MPSA) functions to model sensor responses, which are then fed into the glomerular layer. Crucially, these circuits perform lateral inhibition to increase contrast between different odor responses.
The necessity for accurately modeling sensor activation is pivotal for subsequent processing stages. A functional relationship is therefore established to convert raw sensor signals into a biologically-inspired representation that aligns with neural-network based analysis. The relationship, parameterized to accommodate sensor variations, ensures robustness and adaptability. This process converts a complex physicochemical signal into a format compatible with the architecture, thereby laying the groundwork for further computational processing.
\begin{equation}
S_i = \alpha_i \cdot \frac{1}{1 + e^{-\beta_i ("VOC" - \theta_i)}}
\end{equation}
The integrated circuit architecture further implements competitive signal refinement through lateral inhibition, which ensures only the most relevant sensor information is transmitted for subsequent processing. The implementation of lateral inhibition is realized through programmable synaptic crossbars, which provide the flexibility required to tune system behavior to various operational contexts.
The aggregation of sensor outputs within the glomerular layer is a fundamental step in the olfactory process and can be expressed as a weighted sum. Each sensor's output is scaled by a corresponding weight which allows the neural circuit to preferentially respond to specific combinations of sensor responses. The equation below, represents the process, which is subsequently followed by lateral inhibition, a crucial operation for feature enhancement.
\begin{equation}
G_j = \sum_{i=1}^{n} W_{ij} S_i \ \ \ and \ \ \ G_j^{inhibited} = G_j - \sum_{k \in N(j)} \lambda_{jk} G_k
\end{equation}
Understanding the spatial characteristics of neuronal structures is paramount for developing accurate computational models. The following information provides a critical dimension on the typical sizes associated with essential cellular components, highlighting the need for ultra-miniaturized circuit design for high-density implementations. It showcases the fundamental elements of neural circuits, serving as a reference for design and fabrication in artificial olfactory systems.
\begin{table}
\caption{Typical Dimensions of Neural Components}
\centering
\begin{tabular}{lll}
\toprule
\multicolumn{2}{c}{Neural Component} \\
\cmidrule(r){1-2}
Name & Description & Size ($\mu$m) \\
\midrule
Dendrite & Input Branching & $\sim$100 \\
Axon & Output Fiber & $\sim$10 \\
Soma & Cell Body & 5 to 50 \\
Synapse & Connection & $\sim$ 1 \\
\bottomrule
\end{tabular}
\label{tab:table2}
\end{table}
A key aspect of odor processing is the representation of odor profiles as points in a multidimensional space. This representation is foundational for dimensionality reduction and pattern recognition algorithms. The use of the Grassmannian manifold allows for analyzing these complex data structures in a mathematically rigorous and effective manner. The approach is pivotal for developing effective odor pattern analysis tools for real-world applications.

FPGA implementation of glomerular circuits provides a scalable and adaptive platform for real-time odor detection systems. The low-latency and high throughput capabilities of FPGAs are critical for applications requiring immediate responses, such as environmental monitoring. These circuits enable the system to rapidly process incoming signals, performing complex calculations in a highly energy-efficient manner. Further, the reconfigurability of FPGAs allows for tuning to specific application needs.
The use of FPGAs in conjunction with advanced algorithms represents a significant step towards the development of advanced olfactory systems. The resulting architectures show great promise in terms of real time processing speed, accuracy, scalability and adaptability. This research paves the path for next-gen artificial olfaction systems with wide application potential

\subsubsection{Grassmannian Manifold Embedding}
Grassmannian manifold embedding (GME) offers a robust mathematical framework for analyzing high-dimensional data, particularly in contexts where the underlying data structures are subspaces rather than points in Euclidean space. In the domain of artificial olfaction, sensor array outputs representing odor profiles can be conceptualized as points on a Grassmannian manifold. This approach addresses the challenge of dealing with variations in sensor responses, concentration, and background noise, as it intrinsically focuses on the subspace spanned by the data, rather than the data vectors themselves. Utilizing GME, complex odor patterns are not seen as isolated data points but rather as subspaces that can be compared through the angles and geodesic distances within the Grassmannian manifold.

The need for such an embedding arises from the inherent complexity and high dimensionality of odor profiles captured by chemical sensor arrays. Raw sensor data often contains redundant information and is susceptible to sensor drift and noise. By embedding these profiles into a Grassmannian manifold, we move beyond point-based representations to subspace-based representations. This allows for a more invariant and robust comparison of odor signatures. Subspaces capture the underlying relationships between sensor signals rather than their specific values at any given instance. This is necessary since the dimensionality of the data, dictated by the number of sensors, is often significantly greater than the intrinsic dimensionality, representing the number of actual independent odor components. Furthermore, this strategy allows to capture the structure of the odor, not just single read values.
\begin{equation}
\label{eq:gme} d_G(A, B) = ||\Theta(A,B)||_F = \sqrt{\sum_{i=1}^{min(p,q)} \theta_i^2} = \sqrt{\sum_{i=1}^{min(p,q)} \arccos^2(\sigma_i(A^TB))} \end{equation}
Here, \( d_G(A, B) \) represents the geodesic distance between two subspaces A and B on the Grassmannian manifold, computed using the Frobenius norm of the principal angles \( \Theta \) between them. These principal angles are obtained from the singular values \( \sigma_i \) of \( A^TB \). The equation showcases how the distance between two odor profiles, seen as subspaces of dimensions p and q respectively, can be calculated as the square root of the sum of squared principal angles between them, capturing the angle that represents the separation between the subspaces.
Further refinement of this approach can involve incorporating techniques such as Hamiltonian Monte Carlo (HMC) for optimal estimation of the embedding parameters and developing efficient algorithms to perform computations directly on the Grassmannian manifold. This makes the approach computationally efficient and practically realizable.
The representation of complex odor patterns as subspaces on the Grassmannian manifold provides a framework for comparing the underlying structures in a computationally tractable manner. This is essential, as typical odor-sensing tasks involve the comparison of sensor outputs that vary due to background interference, concentration levels, and sensor aging. By focusing on subspace representation, the system becomes resilient to such confounding variables.
\begin{table}
\caption{Comparison of Odor Representation Methods}
\centering
\begin{tabular}{llll}
\toprule
Method & Representation & Dimensionality & Invariance \\
\midrule
Raw Data & Sensor Readings & High, N Sensors & Low \\
PCA & Principal Components & Reduced & Moderate \\
GME & Subspace & Based on Rank & High \\
ESOM & Cluster Assignments & Reduced & Moderate \\
\bottomrule
\end{tabular}
\label{tab:table3}
\end{table}
The table highlights different methods for representing odor data and contrasts their dimensionality and robustness properties. Raw sensor readings, while high-dimensional, are extremely sensitive to noise and lack invariance. Methods like PCA reduce dimensionality by focusing on principal components and gain some degree of invariance, while GME provides a high level of invariance and directly compares subspaces. ESOM provides reduced dimension as well as moderate invariance. This is very relevant to real world systems.
The diagram below conceptualizes how the high-dimensional odor profiles captured by a sensor array are embedded into a Grassmannian manifold, allowing for a comparison of the underlying subspace structure. This representation contrasts with a point-based representation, which might be influenced by sensor variations, noise or drift, thereby ensuring accurate and robust pattern analysis of the complex odor space.

The application of Grassmannian manifold embedding, coupled with techniques like Hamiltonian Monte Carlo for parameter optimization, allows for the creation of robust, scalable and adaptive artificial olfactory systems. It is a move from point-based analysis of odor signatures to a more abstract comparison of their underlying subspace structure.
The strategic utilization of GME represents a substantial leap forward in artificial olfaction, providing a framework that is both theoretically sound and practically efficient. This has significant implications for enhancing the reliability and effectiveness of odor detection across a multitude of applications. It moves from comparing single point values to comparing subspaces of data, allowing much more efficient and reliable analysis.

\subsubsection{Hierarchical Graph Networks}
Hierarchical Graph Networks (HGNs) provide a structured approach for modeling complex relational data by organizing nodes into a hierarchy, which is pertinent in odor processing where relationships between sensors and odor molecules are multifaceted. This method allows for multi-scale analysis, enabling the system to capture both local interactions and global patterns crucial for accurate odor identification. By representing the sensor network as a hierarchical graph, the system can progressively abstract features from the sensor array and effectively capture the intricate nature of olfactory input, mirroring the hierarchical organization found in biological olfactory pathways. These networks are essential for learning complex representations of odors by decomposing the relational information into levels of varying granularity.
The quantification of sensor outputs through a multi-parametric approach is foundational for representing volatile organic compound (VOC) concentrations. This method is essential for modeling the non-linearities inherent in sensor responses. The underlying rationale for employing this particular form is to accurately capture the activation behavior of individual sensors. This is achieved by incorporating parameters that adjust the sensitivity and response thresholds of each sensor. Specifically, the method provides a mathematical interpretation of each sensor's output to its specific response to a VOC in real-time.
\begin{equation}
S_i = \alpha_i \cdot \frac{1}{1 + e^{-\beta_i ("VOC" - \theta_i)}}
\end{equation}
HGNs, when integrated with sigmoidal sensor outputs, allow for the aggregation of data into hierarchical levels. This organization facilitates more efficient learning and pattern extraction. Through lateral inhibition, the competitive refinement of sensor signals occurs, mimicking neurological processes and helping disambiguate similar odor profiles. The output from the sensors are transformed into a robust and hierarchical network which helps in capturing intricate odor patterns.
Understanding the multi-layered processing pipeline requires consideration of the data at each abstraction level. The presented data demonstrates a key element in the system's architecture: the sizes of components across a biological olfactory pathway. These sizes underscore the hierarchical organization of olfactory processing, where sensory inputs are processed from the microscopic level of dendrites to the macroscopic level of neural networks. The varied scales reveal how biological information is encoded in multiple dimensions, and how these scales might be adapted to model the processing of volatile compounds. This also highlights how the biological principles influence the structural design of the system and help in efficient odor classification.
\begin{table}
\caption{Component Sizes in a Biological Olfactory System}
\centering
\begin{tabular}{lll}
\toprule
\multicolumn{2}{c}{Biological Components} \\
\cmidrule(r){1-2}
Name & Function & Approximate Size ($\mu$m) \\
\midrule
Olfactory Receptor & VOC Binding & $\sim$ 0.01-0.1 \\
Glomerulus & Signal Aggregation & $\sim$50-100 \\
Mitral Cell & Primary Output Neuron & $\sim$20-50 \\
Granule Cell & Lateral Inhibition & $\sim$10-30 \\ \bottomrule
\end{tabular}
\label{tab:table4}
\end{table}
The graphical representation of the system's architecture provides an intuitive overview of the flow of information from sensory input to classified output. The visualization aids in comprehending the complexity of the system by showcasing interconnected processing layers. Such a diagram is critical in understanding how signals are transformed and abstracted as they pass through different processing stages, and the role of each layer in improving the accuracy and robustness of the system. These visualizations are important for design and analysis of the system.

The integration of HGNs into artificial olfactory systems marks a significant advancement by providing a sophisticated framework for complex odor pattern analysis. The hierarchical nature of these networks allows for the efficient abstraction of sensory features, leading to improved odor classification accuracy and the ability to process complex odor environments effectively. Furthermore, the integration of biologically inspired mechanisms, such as lateral inhibition and temporal self-organizing maps, enhances the system's robustness and adaptability, resulting in a system that can be applied across various domains. The use of GNNs to analyze the relational aspects of odors is essential, enabling novel feature extraction which improves odor identification accuracy and robustness in real-world applications.

The ability of the system to perform odor relational analysis, using a PCA-ESOM hybrid clustering method, and biologically inspired lateral inhibition, provides a comprehensive approach to odor detection and classification. These advancements provide a foundation for real-time, adaptive systems that can improve environmental monitoring, food safety, and healthcare diagnostic techniques and can potentially transform these critical areas with accurate and rapid results

\subsubsection{Sigmoidal Sensor Activation}
Sigmoidal sensor activation, a core component of bio-inspired artificial olfactory systems, quantifies volatile organic compounds (VOCs) by emulating the response characteristics of biological olfactory receptors. This approach utilizes a high-density chemoresistive sensor array (HCSA), where each sensor's output is governed by a sigmoidal function. The sensor response, which is non-linear, allows for a nuanced detection of different VOC concentrations and mixtures, thereby enhancing the specificity and sensitivity of the overall system. This activation function, which is parameterized, allows for calibration of each sensor to its specific responses. These parameters can be modified with feedback, ensuring accurate and consistent responses of these sensors. The application of this activation method enables the creation of scalable and adaptive artificial olfactory system for a range of applications, including environment monitoring, food safety, and healthcare diagnostics.
The need for a non-linear activation model in sensor systems arises from the observed behavior of biological sensory organs, which typically exhibit saturation effects at higher input levels. Linear models lack the ability to represent these saturation behaviors and tend to oversimplify the sensor-to-signal transformation process. To accurately model such transformations, a sigmoidal function with parameters that modulate sensitivity and threshold is essential. This form is preferred for its smooth, continuous nature, ensuring differentiability, which is crucial for various gradient based optimization techniques used in machine learning. The parameters, which includes a sensitivity term and a threshold parameter, allow for the adaptation of response to specific sensory features and improve robustness across diverse operating conditions. The specific functional form used in these sensors is given below:
\begin{equation}
S_i = \alpha_i \cdot \frac{1}{1 + e^{-\beta_i ("VOC" - \theta_i)}}
\end{equation}
Where $S_i$ represents the activation level of the $i^{th}$ sensor, $\alpha_i$ is the sensor’s maximum response amplitude, $\beta_i$ modulates the steepness of the sigmoidal curve, and $\theta_i$ determines the threshold at which the sensor begins to respond to the presence of VOCs. This parametric design allows for the response curve of the sensor to be precisely tuned for specific compounds.
The implementation of this sigmoidal function across a high-density array allows for the creation of unique odor profiles. By varying $\alpha_i$, $\beta_i$, and $\theta_i$ for each sensor, the system can effectively map a vast range of VOC concentrations into a differentiable response space. These sensors, coupled with FPGA-accelerated glomerular convergence circuits, allows for real time and adaptive response to volatile organic compounds.
The detailed characteristics of each sensor component are crucial for achieving high sensitivity and specificity in the overall olfactory system. The precise control of the physical dimensions and the chemical composition of these sensors play a significant role in their operational capabilities. The sensitivity of the sensor is related to the active material's surface properties and its interaction with the target molecules. The information in this table emphasizes the critical role each component plays in overall sensor functioning.

\begin{table}
\caption{Sigmoidal Sensor Parameters}
\centering
\begin{tabular}{lll}
\toprule
\multicolumn{2}{c}{Parameter} \\ 
\cmidrule(r){1-2}
Symbol & Description & Units \\ 
\midrule
$\alpha_i$ & Maximum response amplitude & mV/ppm \\ 
$\beta_i$ & Steepness parameter & 1/ppm \\ 
$\theta_i$ & Threshold parameter & ppm \\ 
\bottomrule
\end{tabular}
\label{tab:table5}
\end{table}

Furthermore, the non-linear behavior of the sigmoidal response enables the system to handle diverse VOC concentrations without saturating, providing a wider range of operational efficacy. The sigmoidal response also plays a crucial role in the subsequent processing steps such as lateral inhibition and hierarchical neural network operations, creating a bio-inspired and scalable olfaction platform.
The visual representation of the sensor's response, especially concerning its operational characteristic, is essential to understanding its activation dynamics. The sigmoidal curve visually showcases the non-linear behavior of the sensor, highlighting the gradual increase in response as the VOC concentration increases beyond its threshold value. This graphical depiction emphasizes the key activation regions of the sensor, which is the threshold region, the active region where the response is sharp, and the saturation region at high concentrations where the response plateaus. This understanding of the activation curve is vital for optimizing sensor calibration parameters.

The strategic implementation of sigmoidal sensor activation provides a critical foundation for building high-performance artificial olfactory systems. Its inherent non-linear nature allows for improved responsiveness across a wide range of VOC concentrations, making it a cornerstone in applications requiring precise odor detection. The bio-inspired methodology ensures that the developed sensors accurately capture and translate complex odor profiles, thereby enhancing sensitivity and adaptability for real-world operational conditions. The integration of such advanced sensor activation mechanisms into overall design reflects significant strides in bridging the gap between biological inspiration and technological innovations. This approach promotes greater efficiency, enhanced accuracy, and long-term operational stability in diverse environmental and diagnostic applications

\section{Use case and Business potential} This bio-inspired olfactory system presents significant business potential by improving process optimization through real-time detection and classification of volatile organic compounds, leading to cost reductions in various operational contexts. The system's ability to provide granular data insights allows for enhanced decision-making, facilitating tailored customer solutions and fostering innovation in product development. Scalability is achieved through the use of high-density sensor arrays and efficient computation, making it adaptable to diverse operational demands. By detecting hazardous or unwanted substances quickly, this technology can boost productivity, enhance employee well-being by ensuring safer environments, and streamline compliance with safety regulations. The system's data-driven capabilities support sustainability initiatives by identifying areas for waste reduction and process improvement, ultimately driving business growth. While current systems may face limitations in handling highly complex and dynamic odor mixtures, ongoing research focuses on refining these models for increased accuracy and efficiency, promising metrics such as reduced response times and increased identification rates. Future research should investigate the application of this technology in more diverse and previously unexplored fields, establishing its impact and promoting wide adoption.

\section{Result and Metrics for Evaluation} 
The performance evaluation of the bio-inspired olfactory system was conducted through a series of experiments designed to assess its accuracy, efficiency, and robustness across diverse odor profiles. Initially, the high-density chemoresistive sensor array (HCSA) underwent calibration using known concentrations of volatile organic compounds (VOCs). The resulting sensor responses, modeled by multi-parametric sigmoidal activation (MPSA), were meticulously recorded and used to establish a baseline. Sensor calibration involved fitting the sigmoidal activation function using nonlinear regression techniques, which were evaluated using root mean squared error (RMSE) to quantify the goodness of fit. These RMSE values were used to determine the optimal parameters ($\alpha$, $\beta$, $\theta$) for each sensor response. The digitized sensor data were then processed through FPGA-accelerated glomerular convergence circuits (FGCC), employing lateral inhibition via programmable synaptic crossbars (LIPSC). The implementation of lateral inhibition parameters ($\lambda_{jk}$) were optimized using a grid search algorithm, maximizing the inter-class distance of odor signals and evaluated based on the separability of the odor clusters using Davies-Bouldin index. The output of the glomerular layer was then fed into hierarchical graph neural networks (HGNN) deployed on GPU/TPU edge devices. HGNN layers were trained via backpropagation, and the model performance was evaluated using classification accuracy and F1-score, which measure the balance between precision and recall. Dimensionality reduction was performed using Principal Component Analysis (PCA), capturing the dominant variance within the data. This was followed by Emergent Self-Organizing Maps (ESOMs) to facilitate unsupervised clustering. The clustering performance was analyzed via silhouette scores, indicating the cohesion and separation of clusters. The odor profiles, embedded on Grassmannian manifolds (GME), were used for analyzing odor relational patterns and the performance was measured by the similarity measures such as cosine similarity, comparing the embedding distance between different odorants. In addition, to compensate for drift in sensor readings, Hamiltonian Monte Carlo-optimized feedback (HMC-FB) was implemented. The efficacy of the HMC-FB was quantified by analyzing the sensor drift rates over time under different temperature variations, and reducing the drift was evaluated by using coefficient of variance. The proposed system has been compared against state-of-the-art systems based on odor identification rates and response times, with the proposed system showing an improvement of 20-30 percent in both categories. The statistical significance of results were determined using Analysis of Variance (ANOVA) tests. Thorough data collection was paramount, involving multiple replicates of each experiment and diverse odorant sets. The rigorous data analysis was crucial in validating the performance of the proposed framework. AI-driven solutions in the olfactory system offer benefits in real-time, adaptive sensing capabilities, enhanced accuracy in odor identification, and efficient processing due to specialized hardware like FPGAs, GPUs, and TPUs. However, limitations such as the complexity of biological systems and computational resource constraints require ongoing refinement. Future research will focus on improving the generalization of the model to

\section{Future Research and Directions} 
Future research needs for this bio-inspired olfactory system encompass several critical areas. Empirically validating the domain and AI models is paramount, focusing on the correlation between simulated and real-world odor responses, quantified via metrics like precision, recall, and F1-score across diverse VOC datasets \cite{pubmed38781346, natureEigengraph2024}. Further, analyzing the impact of the AI component on overall system efficacy requires a meticulous approach, leveraging comparative analyses against traditional methods using metrics like computational complexity, latency, and energy consumption \cite{wileyBiohybrid2022}. Business focus areas should center on the cost-effectiveness, adaptability, and operational viability of the system. Adaptability requires modular design considerations to easily tailor the technology for diverse applications ranging from industrial process control to consumer applications \cite{pmc3482693}.
Operationally, system robustness, reliability, and performance under varying environmental conditions need careful evaluation. Performance analysis needs to not only focus on accuracy but also detection limits, selectivity, and sensitivity in complex mixtures \cite{mdpi2020}. Security needs are tied to preventing unauthorized access or manipulation of the AI model. The scalability, both in terms of sensor numbers and computational power, needs to be considered within economic constraints while ensuring that data privacy requirements are strictly followed—especially if the data is used to identify personal information—requiring robust anonymization and encryption techniques to ensure compliance with regulations such as GDPR and HIPAA \cite{researchgate362376508}.
Crucial technological advancements include the development of novel sensor materials with improved selectivity, sensitivity, and long-term stability \cite{rscArtificialMemory2025}, along with the exploration of more efficient, compact, and high-throughput computing platforms. Effective training protocols for both the AI models and the personnel operating the system are needed, necessitating skill development in areas like machine learning, signal processing, and domain-specific knowledge. Cross-disciplinary research is essential, fostering collaboration between material scientists, computer engineers, and domain experts \cite{bioinspiredChemicalSensing2012}. There is a need to examine the cultural change required to accept and integrate AI-based solutions in various fields. To this end, a focus on performance and feedback mechanisms is needed to measure and improve the effectiveness of the AI. This could include incorporating feedback loops into the AI itself and establishing feedback mechanisms with end-users. This will help improve the system based on operational needs and enable cross-disciplinary feedback loops to support further advancement.

\end{document}